\documentstyle[11pt]{article}

\newcommand{\gsim}{\stackrel{>}{\scriptstyle \sim}}
\newcommand{\lsim}{\stackrel{<}{\scriptstyle \sim}}
\newcommand{\gtrsim}{\raisebox{-2pt}{\,\mbox{$\gsim$}\,}}
\newcommand{\lesssim}{\raisebox{-2pt}{\,\mbox{$\lsim$}\,}}
\newcommand{\lll}{<\!\!<\!\!<} % \! = negative thin space

\newcommand{\chap}{null}  % chapter name, not needed here
\newcommand{\mylabel}[1]{\label{\chap.#1}}

\newcommand{\eq}[1]{equation~(\ref{\chap .#1})}
\newcommand{\eqs}[2]{equations~(\ref{\chap .#1}) and~(\ref{\chap .#2})}
\newcommand{\mycaption}[2]{{\footnotesize \caption[#1]{#2}}}

\newcommand{\Fig}[1]{Figure~\ref{\chap.#1}}

\newcommand{\mpl}{m_p}

\newcommand{\ystar}{y_*}
\newcommand{\yend}{y_{end}}

\begin{document}

\thispagestyle{empty}%

{\raggedleft
Preprint: NZ-CAN-RE-94/1 \par
}

\vspace{2cm}

	\begin{center}
	{\Large\bf\expandafter{Inflationary  Perturbations\\
                 and\\
             Exponential Potentials}\par}
	\end{center}
	\vfill
	\begin{center}
		 {\Large Richard Easther \\}
                  \vspace{5mm}
                 {\large
                   Department of Physics and Astronomy \\
                   University of Canterbury \\
                   Private Bag 4800 \\
                   Christchurch \\
                   New Zealand. \\}
        \end{center}
        \vspace{5mm}
        \begin{center}
                   email: r.easther@phys.canterbury.ac.nz \\
	\end{center}

\vfill

\section*{Abstract}

We consider the spectrum of primordial  fluctuations produced by
inflationary models where the inflaton potential is the  sum of
two exponential terms. A wide range of spectra result, with the only
constraint being that the scalar spectrum must have more tilt than the tensor
spectrum. This model can mimic the spectra of most
inflationary models, as well as producing  combinations of scalar and
tensor perturbations not found in common varieties of
slow-rolling inflation.

\vfill
\newpage
\setcounter{page}{1}

\section{Introduction}

Inflationary models based on exponential potentials have received wide
attention. Potentials with exponential terms are associated with
superstrings
\cite{CallanET1986a,CampbellET1991a,FradkinET1985a}, supergravity
\cite{SalamET1984a,Halliwell1987a,CardosoET1992a} and theories with an
extended gravitational sector
\cite{MagnanoET1993a,MignemiET1992a,Wands1993a}. The generic feature of
these models is their use of a conformal transformation
\cite{KalaraET1990a} to move into the Einstein frame, which typically
introduces a scalar field with an exponential potential.

 A potential consisting of a single exponential term produces power-law
inflation \cite{AbbottET1984a,LucchinET1985a,LucchinET1985b,Ratra1992a}
which is interesting both as an inflationary model in its own right, and
because the inflationary growth \cite{SalopekET1990a} and perturbation
spectra \cite{LucchinET1985a,StewartET1993a} are known exactly.  In many
instances though, the potential, $V(\phi)$, may contain more than one
exponential term and in this letter we examine models based upon
\begin{equation}
V(\phi) = A\exp{\left[-\xi\sqrt{\frac{8\pi}{\mpl^2}} \phi\right]} -
   B\exp{\left[-\xi m \sqrt{\frac{8\pi}{\mpl^2}} \phi\right]},
\mylabel{Vphi}
\end{equation}
with $A,B,\xi > 0$ and $m>1$. Any specific theory will predict some or all of
these parameters. However, by considering inflation driven by a potential
with the generic form of \eq{Vphi} we can investigate the cosmological
constraints upon the model parameters.

The potential, \eq{Vphi}, produces a rich variety of inflationary
behaviour. The
essential feature of the model is that $V(\phi)$ has a maximum, near
which it is approximately flat, but is dominated by a single exponential
term at large $\phi$. The two term potential discussed in this
letter is recommended by its simplicity but  a large number of models
have similar observational consequences. For instance, the potential
$V(\phi) \propto \cosh{(c\phi)}^{-2}$ has a maximum at $\phi=0$ and
reduces to a single exponential at large $\phi$. Copeland {\em et.
al.\/} \cite{CopelandET1993a} discuss the inflationary perturbations
produced by this potential, which resemble those found here for
\eq{Vphi}. Exact solutions for potentials consisting of three exponential
terms have been presented by the author \cite{Easther1993b} and the
perturbation spectra for these models are also similar to that produced
by the two term potential, \eq{Vphi}.\footnote{The spectra for the
solutions in \cite{Easther1993b} can be found by the methods used in
this letter.}

As in power-law and intermediate
\cite{Barrow1990a,BarrowET1990c,BarrowET1993b} inflation
 there is no natural end to the inflationary era in this model. We adopt
the usual approach of specifying a value of $\phi$ at which inflation
ends. The mechanism that ends inflation must not produce unacceptably
large inhomogeneities in the microwave background, and must reheat the
universe to a high enough temperature to allow baryogenesis.  For large
negative values of $\phi$ the potential is not bounded below, so we
assume that $\phi$ does not enter this region. Physically, an unbounded
potential often signals the breakdown of some perturbation expansion.

\section{The Perturbation Spectrum}

The Einstein field equations applied to a spatially flat, homogeneous and
isotropic universe dominated by a scalar field give
\cite{KolbBK1,LindeBK1}
\begin{eqnarray}
&& H^2 = \frac{8\pi}{3\mpl^2}
    \left( \frac{\dot{\phi}^2}{2} + V(\phi)\right), \mylabel{Hsqrd} \\
&& \ddot{\phi } + 3H\dot{\phi} + V(\phi)' = 0. \mylabel{eofmotion}
\end{eqnarray}
The Hubble parameter, $H = \dot{a}/a$, where $a(t)$ is the scale factor.
The dot and dash denote differentiation with respect to the time, $t$,
and the field, $\phi$, respectively. In the slow rolling approximation
these equations reduce to
\begin{eqnarray}
H^2 &=& \frac{8\pi}{3\mpl^2}V(\phi), \mylabel{Hsqrtsr} \\
\dot{\phi} &=& -\frac{V(\phi)'}{3H}. \mylabel{phidotsr}
\end{eqnarray}

It will be useful to make the  change of variables,
\begin{eqnarray}
C &=& \frac{B}{A}, \\
 \frac{1-y}{mC}  &=&
  \exp{\left[-(m-1)\xi\sqrt{\frac{8\pi}{\mpl^2}}\phi\right]}.
         \mylabel{ydef}
\end{eqnarray}
The potential, \eq{Vphi}, has its global maximum at $y=0$, and $y=1$
corresponds to $\phi \rightarrow \infty$. We restrict attention to
models where $\phi$ is monotonically increasing and $y\geq 0$.

Adopting the nomenclature of Liddle and Lyth \cite{LiddleET1992b}, the
perturbation spectra are described in terms of
$\epsilon$ and  $\eta$, where
\begin{eqnarray}
\epsilon &=& \frac{\mpl^2}{16\pi}
\left[\frac{V(\phi)'}{V(\phi)}\right]^2, \mylabel{epsilondef} \\
\eta &=& \frac{\mpl^2}{8\pi}\frac{V(\phi)''}{V(\phi)}.
\mylabel{etadef}
\end{eqnarray}
The scalar (density) perturbation index, $n$, and the
tensor (gravitational) perturbation index, $n_g$, are
\begin{eqnarray}
n &=& 1+ 2\eta - 6\epsilon, \mylabel{ndef}\\
n_g &=& -2\epsilon. \mylabel{ngdef}
\end{eqnarray}
It will be useful to define $n$ in terms of $n_d$, where $n_d = n-1$. If
$n_d$ and $n_g$ are zero, then the corresponding spectrum is
flat, or scale-free. Conversely, if the indices are non-zero, then the
spectra are tilted. The ratio of the tensor and scalar perturbation
amplitudes, $R$, is
\begin{equation}
R = 12.4\epsilon.   \mylabel{Rdef}
\end{equation}
The above relationships apply to all inflationary models driven by a
slowly rolling scalar field. From these expressions alone we see that
the scalar and tensor perturbation spectra produced by inflation cannot
be specified arbitrarily. In particular, if the spectrum of primordial
gravitational waves is flat ($\epsilon \lll 1$), then it will also be
of negligible amplitude compared to the scalar
perturbations.

In terms of $y$, the parameters $\epsilon$ and $\eta$ derived from
$V(\phi)$, \eq{Vphi}, are
\begin{eqnarray}
\epsilon &=& \frac{m^2\xi^2}{2}\left[\frac{y}{m-1+y}\right]^2,
               \mylabel{epsilony} \\
\eta &=& m\xi^2 \frac{1 - m + my}{m - 1 + y},
               \mylabel{etay}
\end{eqnarray}
leading to
\begin{eqnarray}
n_d &=& \frac{m\xi^2}{(m-1+y)^2}
             \left[ 2(1-m)^2(y-1) - my^2 \right], \mylabel{ndy} \\
n_g  &=& -m^2\xi^2\left[\frac{y}{m-1+y}\right]^2. \mylabel{ngy}
\end{eqnarray}
These expressions depend only on $m$ and $\xi$, and the field
value $\phi$ (or $y$) at which the perturbations are produced.

The indices $n_d$ and $n_g$ may have a wide variety of values. Since
$\epsilon \geq 0$ by definition, all slowly rolling inflationary models
predict $n_g \leq 0$.  From \eqs{ndy}{ngy} we deduce $n_d \leq n_g$, with
the equality holding only when $y=1$. Hence for this model we have the
relationship
\begin{equation}
n_d \leq n_g \leq 0. \mylabel{nconstraint}
\end{equation}
This constraint is satisfied by almost all inflationary models, with the
exceptions of false vacuum or hybrid inflation
\cite{CopelandET1994a,Linde1993a} and intermediate inflation
 which may have $n_d > 0$. Thus inflation driven by the potential, $V(\phi)$,
can mimic most of the usual models of inflation (tabulated by Liddle and
Lyth~\cite{LiddleET1992b}, for instance). Furthermore, many
combinations of $n_g$ and $n_d$ satisfying \eq{nconstraint} are not
found in common inflationary models.

Consider the limiting cases $y \approx 0$ and $y \approx 1$
separately. When $y\approx 0$, the field point is rolling off the global
maximum of the potential. In this case the tensor perturbations are of
negligible amplitude, since $\epsilon \approx 0$, but the scalar
perturbations may be sharply tilted, with
\begin{equation}
n_d(y=0) = -2m\xi^2.  \mylabel{ndy0}
\end{equation}
This spectrum is identical to that produced by natural inflation
\cite{FreeseET1990b,AdamsET1993a}, where
the field rolls off the maximum of a potential, $V(\phi) \propto 1 -
\cos{(\phi/f)}$.

When $y\approx 1$, $\phi$ is large and the potential is dominated by the
first term in \eq{Vphi}, since it is assumed that $m > 1$. This limiting
case then reduces to power-law inflation driven by a single exponential
potential. The  perturbation spectra for power-law inflation
are known exactly \cite{LucchinET1985a,StewartET1993a} but if the
slow-rolling approximation is valid then
\begin{equation}
n_d(y=1) = n_g(y=1) = - \xi^2. \mylabel{ndy1}
\end{equation}

\section{Observational Constraints}

The strongest observational constraints on inflationary models are
derived from the primordial perturbation spectrum. At large scales, COBE
\cite{SmootET1992a} and the Tenerife experiment \cite{HancockET1994a}
have been combined to give the $1\sigma$ lower bound, $n\geq 0.9$,
although this bound will weaken considerably at the $2\sigma$ level. Adams
{\it et. al.\/} \cite{AdamsET1993a} review the limits on $n$ derived from a
variety of sources. They conclude that the COBE results, combined with
the requirement that galaxies form sufficiently early, requires $n \geq
0.6-0.7$, and that these bounds are raised if tensor perturbations make
a significant contribution to the COBE signal. On the other hand, the
APM survey \cite{MaddoxET1990a} together with standard Cold Dark Matter
(CDM) requires a sharply tilted scalar index, with $n\leq0.6$. Liddle
and Lyth \cite{LiddleET1992c} combine the COBE data with the QDOT
survey \cite{SaundersET1991a} to obtain $n>0.7$ for a purely scalar
perturbation, rising to $n>0.84$ for power-law inflation, which has a
significant contribution from tensor perturbations.

Clearly, not all of these bounds can be satisfied simultaneously.
However, constraints at shorter scales are sensitive  to the composition
of the dark matter, the baryon content of the universe, a non-zero
cosmological constant and scale dependent biasing. Microwave background
experiments at angular scales above one degree, such as COBE and Tenerife,
provide a direct probe of the primordial spectrum since these scales
have not evolved significantly since they re-entered the horizon.
Consequently, we will assume that the lower bounds on $n$ described above
are more robust, and fit the model discussed in this letter to them.
This is a more stringent constraint than fitting to the APM
data, since the model here has no difficulty producing large amounts of
tilt.

For definiteness we take the bounds derived by Liddle and Lyth
\cite{LiddleET1992c}. They give the following constraints from
QDOT (with CDM) and the One Year COBE DMR anisotropy, respectively:
\begin{eqnarray}
0.8\left(\frac{2.9 - n}{1.9}\right)^{2/3} < &b_8&
   < 1.4\left(\frac{2.9 - n}{1.9}\right)^{2/3}  \mylabel{b8iras}\\
0.7\sqrt{1+R}\exp{[2.62(1-n)]} < &b_8&
   < 1.5\sqrt{1+R}\exp{[2.62(1-n)]} \mylabel{b8cobe}
\end{eqnarray}
where $b_8$ is the bias factor and the uncertainties are taken at the
$2\sigma$ level. The COBE limit increases if there is a significant
component of tensor perturbations, whose relative amplitude is
given by $R$.

Treating $\ystar$, the value of value of $y$ at which presently
observable perturbations are generated, as adjustable, the spectra are
determined by $\xi$, $m$ and $\ystar$. The region of parameter space
where the constraints of \eqs{b8iras}{b8cobe} are both satisfied is
plotted in \Fig{constraint}. While the numerical bounds adopted here are
generous, the shape of the allowed region of parameter space is
relatively insensitive to them but shrinks as the bounds are tightened.

The limiting values of $\ystar=0$ and $\ystar=1$ correspond to the
spectra produced by natural and power-law inflation, respectively. For
$\ystar=0$, the constraints on $b_8$ require $n \geq 0.7$, or $2m\xi^2
\leq 0.3$. When $\ystar=1$ the added contribution from tensor
perturbations tightens this limit to $n\geq0.85$, or $\xi^2 \leq 0.15$.
If $m\xi^2 \leq 0.15$ (remembering $m>1$) the resulting spectrum
satisfies the constraint of \eqs{b8iras}{b8cobe} at all values of
$\ystar$.   Conversely, if $\xi \gtrsim 0.5$ there
are no values of $m$ and $\xi$ which produce a viable perturbation
spectrum. However, while $\xi \lesssim 0.5$, the
coefficient in the second exponential in $V(\phi)$, $m\xi$, may be of
order unity.

The dominant second order corrections  to the spectral indices
\cite{StewartET1993a} are proportional to $m^2\xi^4$. If $m \lesssim 5$,
then the first order expressions for $n$ and $n_g$, \eqs{ndef}{ngdef},
are valid whenever the inequalities of \eqs{b8iras}{b8cobe} are satisfied.  At
larger values of $m$ the second order corrections may be large, even
when the spectral indices calculated from the first order expressions satisfy
the constraints from COBE and QDOT.

Since $\xi \lesssim 0.5$ to satisfy the constraints on the
perturbations, the model has no difficulty providing sufficient
inflation to solve the usual cosmological problems.  For given values of
$m$ and $\xi$, the perturbation spectra seen in the sky at the present
epoch depend on the value of $\phi$ (or, equivalently, $y$) at which
inflation ceases, and less strongly on the exact mechanism that brings
about the transition from inflationary to ordinary expansion. The
matching equation \cite{Turner1993a} then determines the value of $\phi$
roughly sixty $e$-foldings before the end of inflation when the
presently observable perturbations were formed. If inflation finishes at
$y=y_{end}$ then $\ystar$ can be calculated from the matching equation
and
\begin{equation}
N_* = \frac{8\pi}{\mpl^2}\int_{\phi_{end}}^{\phi_*}
     {\frac{V}{V'}d\phi} ,
\end{equation}
where $N$ is the number of $e$-foldings before inflation ceases. In terms of
$y$,
\begin{equation}
\frac{(\yend - 1)^m}{\yend^{m-1}}\exp{[m(m-1)\xi^2 N_*]}
   = \frac{(\ystar - 1)^m}{\ystar^{m-1}}.
\end{equation}
If $\yend$ is not close to $1$ and $m$ is large, then $\ystar \ll 1$,
corresponding to the field point starting at a value of $\phi$ close to
the maximum of the potential, $V(\phi)$. However, for smaller values of
$m$ or $\xi \ll 1$, this condition implies $\ystar$ and $\yend$ have
similar values. While this constraint is expressed in terms of $y$,
physical constraints are applied to $\phi$. From \eq{ydef}, we see that
a condition on $y$ will be satisfied over a larger range of $\phi$ if
$\xi$ or $(m-1)$ is small, weakening any fine-tuning problems.

Finally, the amplitude of the perturbations
will fix one of the parameters $A$ and $B$, or alternatively $A$ and
$C$. From the COBE detection, we have the final constraint \cite{LiddleET1992b}
\begin{equation}
\left[\frac{V(\phi_*)}{\epsilon(\phi_*)}\right]^{1/4}
     = 6.2\times 10^{16} \mbox{GeV}.
\end{equation}
which will fix $A$ in terms of $C$, $\ystar$, $m$ and $\xi$. If
gravitational waves make a significant contribution then this value will
be lowered.

\section{Discussion}

We have seen that slow-rolling inflation driven by a potential,
$V(\phi)$, that consists of two exponential terms
produces a wide variety of inflationary perturbations. The versatility
of this model is a consequence of the different inflationary regimes
near the maximum of $V(\phi)$ and at large values of $\phi$. The
observed perturbation spectrum depends on the portion of the potential
the field ``saw'' when the perturbations that seeded the presently
observable structure in the universe were formed. This is determined by
the value of $\phi$ at which inflation ceases.

Clearly, the presence of several free parameters in $V(\phi)$ is
necessary for an inflationary model to be able to produce a wide range
of values of $n$ and $n_g$. However, this is not a sufficient condition.
Changing the sign of $B$ in \eq{Vphi} does not alter the number of free
parameters, but the resulting perturbation spectrum is always
approximately similar to that found in a power-law model.

Recently, there has been considerable interest in reconstructing the
potential of a slowly rolling scalar field that is hypothesised to have
produced the primordial perturbation spectrum
\cite{CopelandET1993a,CopelandET1993b,CopelandET1993c}.  Since the model
developed in this letter can mimic most varieties of
inflation, it underlines the point that reconstructions of
inflation do not unambiguously identify the inflaton potential, as
they do not sample a large enough piece of $V(\phi)$, especially while
experimental data is not sufficiently sensitive to probe the scale
dependence of $n$ and $n_g$.

Inflation driven by a potential composed of two exponential terms, such
as \eq{Vphi}, produces combinations of scalar and tensor perturbations
not found in any common slow-rolling model.  In particular, a tilted
spectrum of scalar perturbations, which is one of the possible solutions
to the excess small scale structure in standard CDM, is predicted by
both natural and power-law inflation.  The tensor perturbations are
either negligible (natural inflation) or have the same tilt as the
scalar perturbations (power-law inflation). The model developed in this
letter may also have a  tilted scalar spectrum. The only
constraint on the tensor spectrum, though, is that it has less tilt than
than the scalar one,  thus broadening the range of possibilities
open to the inflationary model builder. Furthermore, exponential
potentials are typically associated with power-law inflation, with its
characteristic perturbation spectrum.  As the model developed
here demonstrates, a potential with more than one exponential term may
produce a broad range of perturbations, significantly increasing the
types of inflationary behaviour that can be associated with exponential
potentials.

\newpage
%\bibliography{massive,books}
%\bibliographystyle{jal}

\newpage

\section*{Figure Caption}

\begin{figure}[htpb]
\begin{center}
\begin{minipage}{10cm}
\mycaption{}{The region of parameter space
where the constraints on $b_8$ from QDOT and COBE are both satisfied by
the inflationary perturbation spectrum produced by $V(\phi)$,
\eq{Vphi}. \mylabel{constraint}}
\end{minipage}
\end{center}
\end{figure}

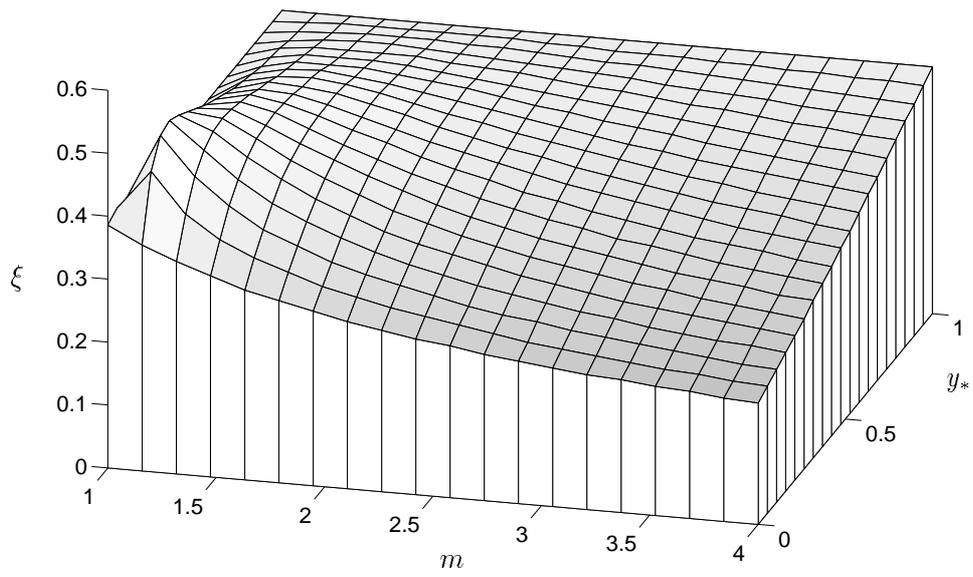
\end{document}